\documentclass[prl,twocolumn,showpacs,floatfix]{revtex4-1}
\usepackage{graphics,epsfig,amsfonts,amssymb,amsmath,soul}
\begin{document}

\title{Glassy features of crystal plasticity}

\author{Arttu Lehtinen$^1$, Giulio Costantini$^2$, Mikko J. Alava$^1$, 
Stefano Zapperi$^{2,1,3,4}$, and Lasse Laurson$^1$}
\affiliation{$^1$COMP Centre of Excellence,
Department of Applied Physics, Aalto University, P.O.Box 11100, 
FI-00076 Aalto, Espoo, Finland.}
\affiliation{$^2$Center for Complexity and Biosystems,
Department of Physics, University of Milano, via Celoria 26, 20133 Milano, Italy}
\affiliation{$^3$ISI Foundation, Via Alassio 11/C, 10126 Torino, Italy}
\affiliation{$^4$CNR-IENI, Via R. Cozzi 53, 20125 Milano, Italy}

\begin{abstract}
Crystal plasticity occurs by deformation bursts due to the avalanche-like motion
of dislocations. Here we perform extensive numerical simulations of a 
three-dimensional dislocation dynamics model under quasistatic 
stress-controlled loading. Our results show that avalanches are power-law 
distributed, and display peculiar stress and sample size dependence: The average 
avalanche size grows exponentially with the applied stress, and the amount of 
slip increases with the system size. These results suggest 
that intermittent deformation processes in crystalline materials exhibit an 
extended critical-like phase in analogy to glassy systems, instead of 
originating from a non-equilibrium phase transition critical point.
\end{abstract}
\maketitle

\section{Introduction}
Plastic deformation of crystalline solids, mediated by the stress-driven 
motion of crystal dislocations, has been shown to be a highly 
heterogeneous and wildly fluctuating process \cite{ZAI-06,ALA-14,ANA-07}, in analogy
to numerous other driven systems exhibiting ``crackling noise'' \cite{SET-01}. Broad, 
power-law like distributions of strain bursts are observed in experiments 
on micron-scale samples \cite{UCH-09,DIM-06,CSI-07,NG-08,BRI-08,ZAI-08,
FRI-12,MAA-15,PAP-12}, and the same is often true for acoustic emission (AE) 
amplitudes in the case of larger specimens \cite{WEI-15,WEI-97,MIG-01}. 
While the bursty nature of crystal plasticity is a well established fact, 
the question of its nature and origin remains a subject of a lively 
debate \cite{ISP-14,FRI-12,MAA-15}.

To address such questions in an appropriate fashion, high quality numerical studies of 
realistic discrete dislocation dynamics (DDD) models, capturing the avalanche-like 
deformation process, are essential \cite{ISP-14,OVA-15,CSI-07,JAN-15}. The majority of 
DDD studies of dislocation avalanches have so far been performed
using relatively simple and computationally efficient 2D systems, describing 
point-like cross-sections of ensembles of straight, parallel edge dislocations 
\cite{ISP-14,OVA-15,JAN-15}. Real three-dimensional plastically deforming crystals are 
not described in all their aspects by the 2D DDD models \cite{KAP-15}. In 3D, 
dislocations are flexible lines (exhibiting in general a mixture of edge and
screw character) gliding along multiple slip planes, and interacting in addition 
to the long-range elastic stress fields also via various short-range dislocation 
reactions \cite{FRE-09} (junction formation, annihilation, etc.). During the deformation 
process dislocation density typically increases due to e.g. growth of dislocation 
loops, and via the activation of Frank-Read sources, thus leading to strain 
hardening of the material. It is tempting to attribute the complexity to an underlying 
phase transition with divergent correlations, so that for high stresses above the yield 
stress continuous flow would ensue. A scaling picture related to mean-field -like 
behavior (due to long-range interactions) and a pinning/depinning transition 
(arising from the mutual interactions among moving and jammed dislocations),
has been proposed \cite{CSI-07,FRI-12,MAA-15}. However, dislocations do not in general 
move in the presence of a static pinning field, and therefore they tend to ``jam'' 
instead of getting pinned; moreover, their mutual interactions are anisotropic and 
non-convex, implying that e.g. the no-passing theorem would not be applicable.

\begin{figure}[t!]
\leavevmode
\includegraphics[clip,width=0.9\columnwidth]{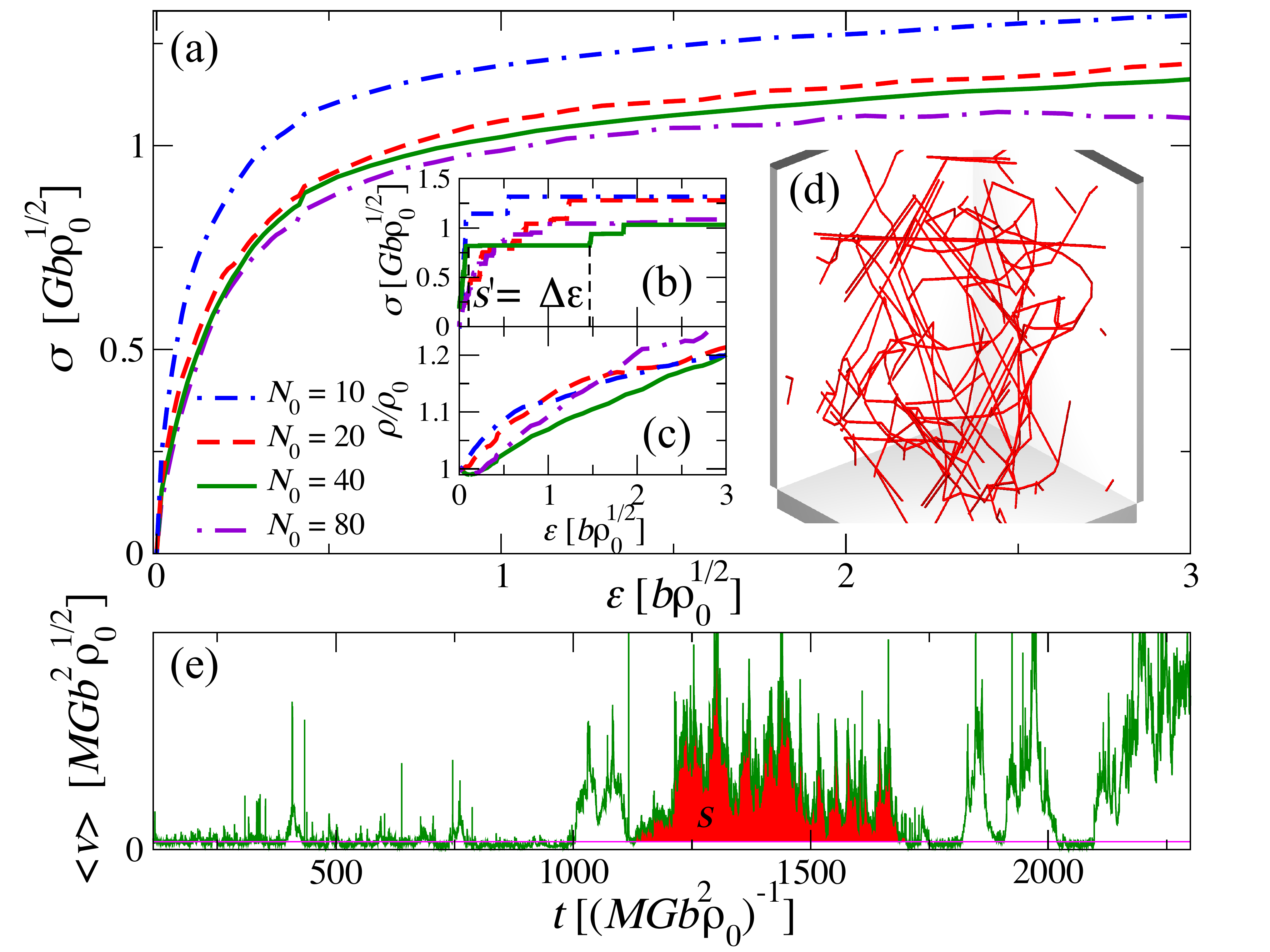}

\caption{(color online) (a) The average stress-strain curves $\sigma(\epsilon)$ for 
different system sizes $N_0$, revealing a size effect (``smaller is stronger''). 
(b) Examples of individual, staircase-like stress-strain curves, also showing 
the definition of a strain burst avalanche size $s' = \Delta \epsilon$.
(c) Average evolution of the dislocation density $\rho$ with $\epsilon$. (d)
An example of the dislocation configuration with $N_0 = 40$, deformed 
up to $\epsilon = 3$ $b\rho^{0.5}$. (e) A typical time series of the
average dislocation velocity $\langle V \rangle (t)$; the red area under a burst
shows an example of the definition of the size $s$ of a velocity avalanche.}
\label{fig:1}
\end{figure}

In this work, we present results from an extensive study of dislocation
avalanches in a fully 3D DDD model. We show that the bursty 3D plastic deformation 
process exhibits scale-free features already at the beginning of the stress-strain 
curve. The small-scale and large-scale (``collective'') avalanches have different 
scaling exponents, and the average avalanche size increases exponentially with 
the applied stress, in analogy with the dynamics observed in simple 2D models 
\cite{ISP-14,JAN-15} which, however, miss completely the strain hardening present 
in 3D. In our DDD simulations of Al single crystals, performed using the ParaDis 
\cite{ARS-07} code, we employ a quasistatic stress-controlled loading protocol. 
This eliminates possible rate effects in the avalanche statistics such as occur
e.g. in the ABBM model of mean field avalanches \cite{ALE-90}. Our detailed 
statistical analysis of the sizes (for durations see Supplemental Material \cite{SM}) of 
the deformation bursts - encompassing both stress-resolved and integrated probability 
distributions - reveals a novel scaling picture which is at odds with the mean-field 
depinning scenario \cite{ZAI-06,CSI-07,FRI-12,MAA-15}. Instead, plasticity of FCC 
single crystals is found to exhibit an extended, critical-like phase, with the amount 
of slip within strain bursts diverging with the system size at any applied stress,
hinting at a system-spanning correlation length. We attribute such behaviour to the 
glassy properties of the dislocation system \cite{ISP-14,BAK-07}, originating from 
frustrated dislocation interactions.

\section{Simulations}

In this work we use a modified version of the DDD code ParaDis  \cite{ARS-07}. In ParaDis dislocations are modeled using a nodal discretization scheme: dislocation lines are 
represented by nodal points connected to their neighbors by straight segments. Changes in dislocation geometry are made possible by adding and removing these nodal points. The total stress acting on a node consists of the external part, resulting from the deformation of the whole crystal, and of the internal, 
anisotropic stress-fields generated by the other dislocations within the crystal. The latter stress fields are computed by applying the well-known results of linear elasticity theory to the straight segments between nodes. Both of these fields generate forces which move the discretization nodes. The external stress generates a Peach-Koehler force which is applied to all nodes. The forces between dislocations themselves are divided to local and far field ones. Forces between segments of nearby nodes and self-interaction of dislocations which are calculated with explicit line integrals. Far-field forces are calculated from the coarse-grained dislocation structure using a multipole expansion. Near the dislocation core, local interactions, such as junction formation, annihilation,  
etc., are introduced phenomenologically with input from smaller scale simulation methods (e.g. MD) and 
experimental results. Once the forces are known, a trapezoidal integrator is used to solve the equation of motion for the discretization nodes. However it must be taken into account that  in real  materials, the motion of dislocations is subject to constraints which depend on the underlying crystal structure (e.g FCC or BCC) and the nature of 
the dislocations (e.g. screw or edge) in a complicated manner. These details are encoded 
in the material-specific mobility function which relates the total forces experienced by 
dislocations to their velocities. In order to simulate bulk properties we use Periodic Boundary Conditions (PBC). These are implemented by using an Ewalds sum procedure similar to those used in atomistic simulations with periodic structures and long range interactions. The main simulation cell is surrounded by periodic images cells which contain the images of the segments in the main cell. Interaction stresses between given segment and its images is obtained from precomputed tables which contain the possible imagestresses from differential segments as function of dislocation orientation and Burgers vector \cite{Bulatov-00}.

We consider here the FCC crystal structure with  
material parameters of Al (shear modulus $G = 26$ GPa, Poisson ratio 0.35,
Young modulus 70.2 GPa, Burgers vector $b = 2.863 \times 10^{-10}$ m, and dislocation 
mobility $10^{4}$ Pa$^{-1}$s$^{-1}$; for simplicity, both edge and 
screw segments are taken to have the same mobility), and employ periodic boundary 
conditions. The typical maximum strains in the simulations are of the order of 1 \%, 
limited by computational cost and what is physically feasible given the boundary 
conditions.
To study the effect of the system size, we consider different linear sizes $L = 0.715$,
1.001, 1.2298, 1.43 and 2.1473 $\mu$m of the cubic simulation box (i.e. within 
the range of those of typical microcrystal compression experiments 
\cite{UCH-09,DIM-06,NG-08,BRI-08,ZAI-08,FRI-12,MAA-15}), keeping 
the initial dislocation density roughly constant at $\rho_0 \approx 3.0 \cdot10^{13}\,$ 
1/m$^2$; this leads to initial numbers $N_0=10$, $20$, $30$, $40$ 
and $80$ of straight mixed dislocations placed randomly on the 
glide planes of the FCC lattice. Results are shown in scaled units, by measuring
lengths, stresses, strains and times in units of $\rho_0^{-1/2}$, $Gb\rho_0^{1/2}$, 
$b\rho_0^{1/2}$, and $(Gb^2M\rho_0)^{-1}$, respectively. For more technical details 
of the 3D DDD simulations, see Supplemental Material \cite{SM}.

The random initial 
configurations are first relaxed in zero applied stress, a process during which 
the dislocation network evolves towards a (meta)stable state where the initially 
straight dislocation lines exhibit some curvature. After relaxation, the quasistatic
stress-controlled driving is initiated; to test the robustness of our results, 
we employ two different driving protocols, and also consider simulations with (Supplemental
Material \cite{SM}) and without (results shown in the main article) cross-slip;
while cross-slip affects the hardening rate (Supplemental
Material \cite{SM}), the strain burst statistics is unaffected by it. The 
dislocation activity is measured either by the absolute collective segment-length 
weighted dislocation velocity $V(t)=(\sum_i l_i  v_{\perp,i})/(\sum_i l_i)$ 
(with $l_i$ and and $v_{\perp,i}$ the length and velocity perpendicular
to the line direction of the $i$th dislocation segment, respectively), or by the strain 
rate $\dot{\epsilon}(t)$ (originating from dislocations moving in the direction 
of the resolved applied shear stress). When the activity falls below a small threshold
$V_\text{th}$, the external stress $\sigma_\text{ext}$ is increased at a constant rate 
(we consider $\dot{\sigma}_\text{ext}=2.5 \times 10^{13}$ Pa/s, or 0.0011268 in the scaled
units, unless stated otherwise). When $V(t)$ [or $\dot{\epsilon}(t)$, 
depending on the protocol used] exceeds the threshold, $\sigma_\text{ext}$ 
is kept constant until the avalanche has finished, and $V(t)$ [$\dot{\epsilon}(t)$] 
again falls below the threshold. Here, we focus on velocity avalanches defined by 
thresholding the $V(t)$ signal, with the avalanche size defined
as $s=\int_0^T [V(t)-V_\text{th}]\text{d}t$ ($T$ is the duration of the avalanche such 
that $V(t)$ continuously exceeds $V_\text{th}$); details 
on other protocols and avalanche definitions (e.g. $s'$ defined as the strain increment
$\Delta \epsilon$), leading to essentially the same results, are provided as Supplemental 
Material \cite{SM}, along with an example animation of the bursty  deformation process. 
Fig. \ref{fig:1} shows examples of the simulated average and individual stress-strain 
curves, the evolution of the dislocation density, a snapshot of a deformed dislocation 
configuration, as well as an example of a $V(t)$ signal, including also an 
illustration of the definition of $s$.

\section{Results}

\begin{figure}[t!]
\leavevmode
\includegraphics[clip,width=0.9\columnwidth]{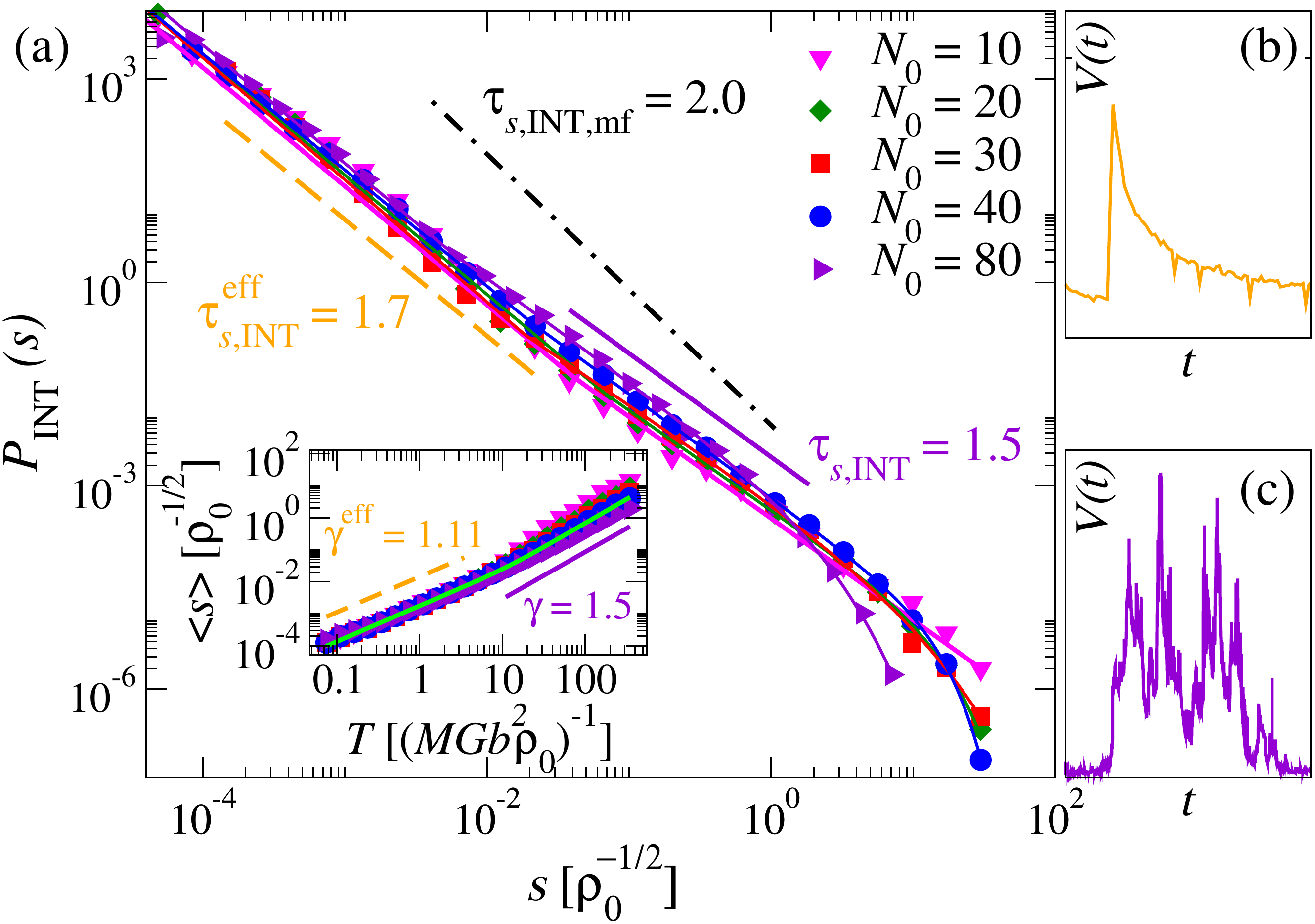}
\caption{(color online) (a) Main panel: stress-integrated avalanche
size distributions $P_\text{INT}(s)$ for different $N_0$.
The data is fitted with the crossover scaling form [Eq. (\ref{eq:xover}), solid lines],
revealing two power law regimes, with small avalanches [an example $V(t)$ signal 
shown in (b)] characterized by $\tau_{s,\text{INT}}^\text{eff} = 1.75 \pm 0.03$,
while larger avalanches [an example $V(t)$ signal shown in (c)] have 
$\tau_{s,\text{INT}}=1.52 \pm 0.04$ (for $N_0 = 80$; considering smaller $N_0$ yields
similar results). For comparison, the dash-dotted line corresponds 
to the mean field result $\tau_{s,\text{INT,MF}}=2$. The inset of (a) displays 
the $\langle s(T) \rangle$ relation, with $\gamma = 1.50 \pm 0.02$ 
and $\gamma^\text{eff} = 1.11 \pm 0.02$ for $T \gg T^*$ and $T \ll T^*$,
respectively.}
\label{fig:2}
\end{figure}

The stress-integrated avalanche size distribution $P_\text{INT}(s)$, i.e. the distribution
of all avalanches irrespective of the $\sigma$-value at which they occur, is the quantity
measured in many experiments \cite{UCH-09,DIM-06,NG-08,BRI-08,ZAI-08,ZHA-12}. 
Fig. \ref{fig:2} shows our $P_\text{INT}(s)$ from the 3D DDD simulations 
for different $N_0$; These exhibit two power law regimes, with a crossover scale 
$s^*$ separating scaling regimes of ``small'' and ``large'' avalanches; analysis 
of the strain burst distributions $P(s')$ (Supplemental Material \cite{SM}) 
shows that $s^*$ corresponds roughly to the characteristic strain burst size 
$s'_1 \propto 1/N_0$, i.e. the strain accumulated due to one dislocation moving one 
average dislocation spacing. Thus, our data is well-described by a crossover
scaling form \cite{LAU-14}
\begin{equation}
P_\text{INT}(s) = \frac{A s^{-\tau_{s,\text{INT}}}}{e^{\left(\frac{s}{s_0}\right)^b}} 
\left[1+\left(\frac{s}{s^*}\right)^{(\tau_{s,\text{INT}}-
\tau_{s,\text{INT}}^\text{eff})\kappa}\right]^{\frac{1}{\kappa}},
\label{eq:xover}
\end{equation}
where $\kappa$ controls the sharpness of the crossover between the two power laws
with exponents $\tau_{s,\text{INT}}$ and $\tau_{s,\text{INT}}^\text{eff}$, and $s_0$
is the cutoff avalanche size, arising here due to the maximum strain reached in the
simulations. We find $\tau_{s,\text{INT}} = 1.52 \pm 0.04$ 
for $s \gg s^*$ in the $N_0=80$ system (similar values are obtained for smaller $N_0$), 
spanning almost three orders of magnitude, while for $s \ll s^*$ a larger 
effective exponent $\tau_{s,\text{INT}}^\text{eff} = 1.75 \pm 0.03$ ensues.
The latter avalanches are small and temporally asymmetric [typically consisting of a 
small jump of an individual dislocation, followed by relaxation, see Fig. \ref{fig:2} 
(b)] like avalanches triggered by local perturbations in a 2D DDD model \cite{JAN-15}. 
Experimental values are scattered around $\tau_{s,\text{INT}} = 1.5$, with some variation 
between different experiments \cite{UCH-09,DIM-06,NG-08,BRI-08,ZAI-08}, in good agreement
with our large-avalanche regime [see also Fig. \ref{fig:2} (c)]; notice that due to 
limited resolution, the small-avalanche regime is not accessible in typical experiments. 

In the inset of Fig. \ref{fig:2}(a), we show the scaling of the 
average avalanche size $\langle s(T) \rangle$ with the avalanche duration $T$;
again two scaling regimes can be observed, and fitting a crossover scaling form
$\langle s(T) \rangle = B T^{\gamma}[1+(T/T^*)^{(\gamma^\text{eff}-\gamma)\kappa}]^{1/\kappa}$
to the $N_0 = 40$ data results in $\gamma^\text{eff} = 1.11 \pm 0.02$ for 
$T \ll T^* \approx 12$, and $\gamma=1.50 \pm 0.02$ for $T \gg T^*$; the latter may be 
contrasted with the mean field depinning value $\gamma_\text{MF}=2$. The large avalanche 
regime has a system size dependent prefactor which can be scaled away by considering an
alternative, ``extensive'' measure of the avalanche size, e.g. the accumulated slip
$\langle d \rangle \equiv \langle s \rangle L^2$ (Supplemental Material \cite{SM}).

\begin{figure}[t!]
\leavevmode
\includegraphics[trim=0.0cm 0.0cm 0.0cm 0.0 cm,clip=true,width=0.9\columnwidth]{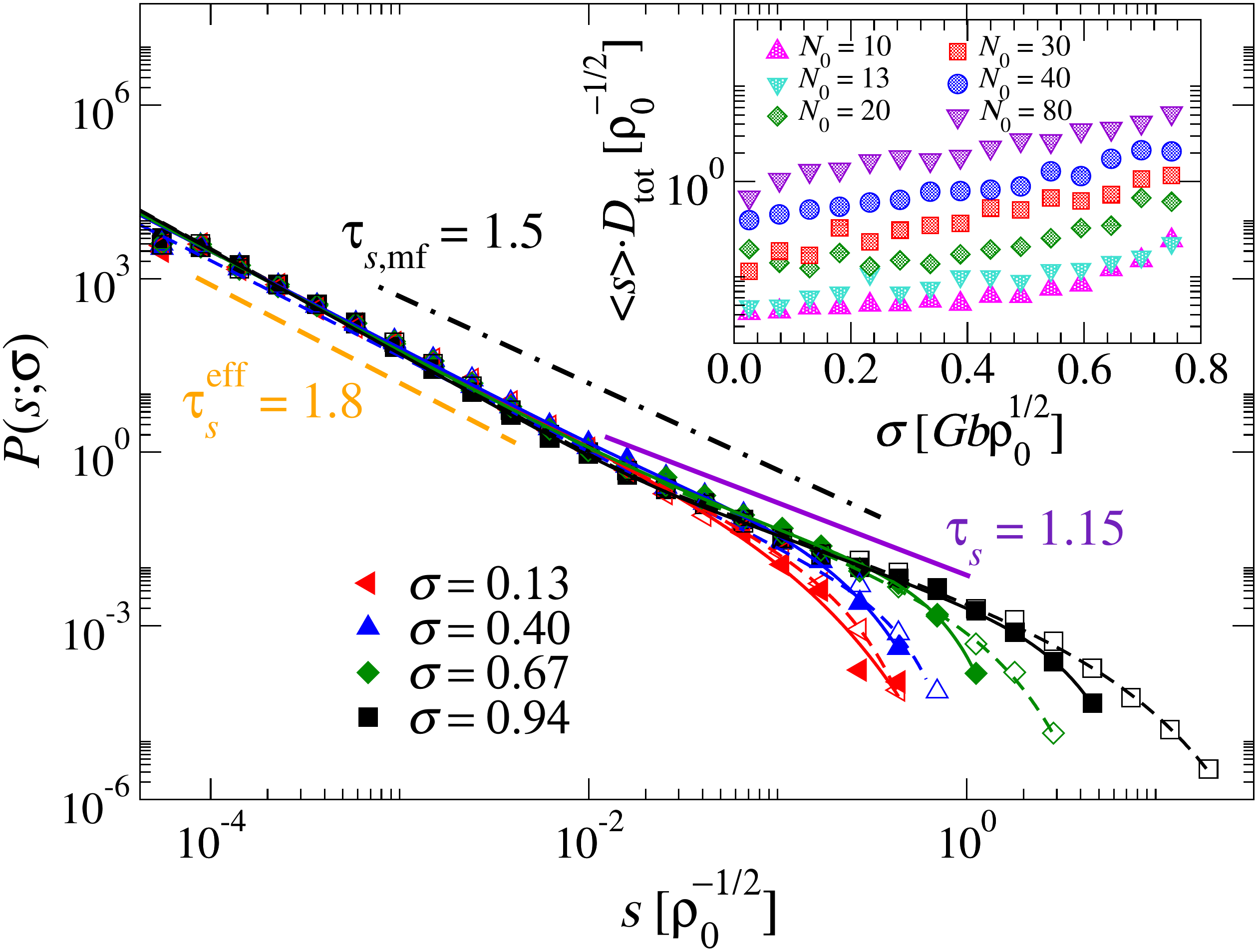}
\caption{(color online) Stress-resolved avalanche size distributions $P(s,\sigma)$ 
for different stress levels $\sigma$ (main figure; open and filled symbols 
correspond to $N_0=40$ and 80, respectively). The data is fitted with the 
crossover scaling form of Eq. (\ref{eq:xover}) (solid lines), revealing 
$\tau_s = 1.18 \pm 0.06$ for $s \gg s^*$, and a larger effective $\tau_{s}^\text{eff} = 
1.80 \pm 0.04$ for $s \ll s^*$ (for $N_0 = 80$; smaller $N_0$'s yield similar values).  
The inset shows the average avalanche size $\langle s \rangle D_\text{tot}$ as a function 
of $\sigma$ for various $N_0$, revealing a roughly exponential $\sigma$-dependence,
and an increasing average avalanche size with $N_0$ at a fixed $\sigma$.}
\label{fig:3}
\end{figure}

It has been proposed that observations of $\tau_{s,\text{INT}} \approx 1.5$ may be
compatible with mean field depinning if, due to back-stresses induced by strain
hardening, the system is constantly pushed towards a critical yield stress, in the
spirit of self-organized criticality (SOC) \cite{CSI-07,ALC-15,BAK-87}, resulting in a stationary 
avalanche process. In Fig. \ref{fig:3}, we consider the stress-resolved avalanche size 
distributions, i.e. $P(s;\sigma)$ of avalanches within stress bins centered at $\sigma$,
as also reported for some experiments \cite{FRI-12,MAA-15}.
Fitting the scaling form of Eq. (\ref{eq:xover}) to the $P(s;\sigma)$ distributions
(with substitutions $P_\text{INT}(s) \rightarrow P(s;\sigma)$, $\tau_{s,\text{INT}} 
\rightarrow \tau_s$, and $\tau_{s,\text{INT}}^\text{eff} \rightarrow \tau_s^\text{eff}$)
reveals a large-avalanche exponent $\tau_s < \tau_{s,\text{INT}}$, a signature of 
non-stationary avalanche processes \cite{DUR-06}; for the largest stress bin in 
Fig. \ref{fig:3}, we obtain $\tau_s = 1.18 \pm 0.06$, while for $s \ll s^*$, a 
larger effective $\tau_{s}^\text{eff} = 1.80 \pm 0.04$ is again observed ($N_0 = 80$; smaller
$N_0$'s yield similar values).
The avalanche cutoff scale $s_0 (\sigma)$ grows with the stress level $\sigma$. This,
together with the fact that the $\tau_s$ exponent is significantly smaller than the 
mean-field \cite{FRI-12} or ABBM \cite{ALE-90} value of 1.5 (notice that we have 
eliminated possible rate effects by employing the quasistatic driving protocol, 
and also verified the indepence of the results on the stress rate, see Supplemental 
Material \cite{SM}), provides strong evidence suggesting that our avalanches cannot be 
described by mean field depinning. The same values for 
$\tau_s$ and $\tau_{s,\text{INT}}$ can be extracted also from the complementary cumulative 
distributions functions (CDFs, see Supplemental Material \cite{SM}), highlighting the 
robustness of the values. 
We also note that these, together with the exponents of the 
duration distributions ($\tau_T = 1.22 \pm 0.14$ and $\tau_T^\text{eff}=1.92 \pm 0.08$, 
Supplemental Material \cite{SM}) fulfill the scaling relations $\gamma=(\tau_T-1)/(\tau_s-1)$ 
and $\gamma^\text{eff}=(\tau_T^\text{eff}-1)/(\tau_s^\text{eff}-1)$ within errorbars 
above and below the crossover, respectively. The values of $\tau_s^\text{eff}$, 
$\tau_T^\text{eff}$ and $\gamma^\text{eff}$ are close to those found recently for 
avalanches triggered by local perturbations in a 2D DDD model \cite{JAN-15}.
Furthermore, our results are not sensitive to details of the preparation of the
initial state, as evicenced by considering systems with a loading history as initial
states (Supplemental Material \cite{SM}).

\begin{figure}[t!]
\leavevmode
\includegraphics[clip,width=0.9\columnwidth]{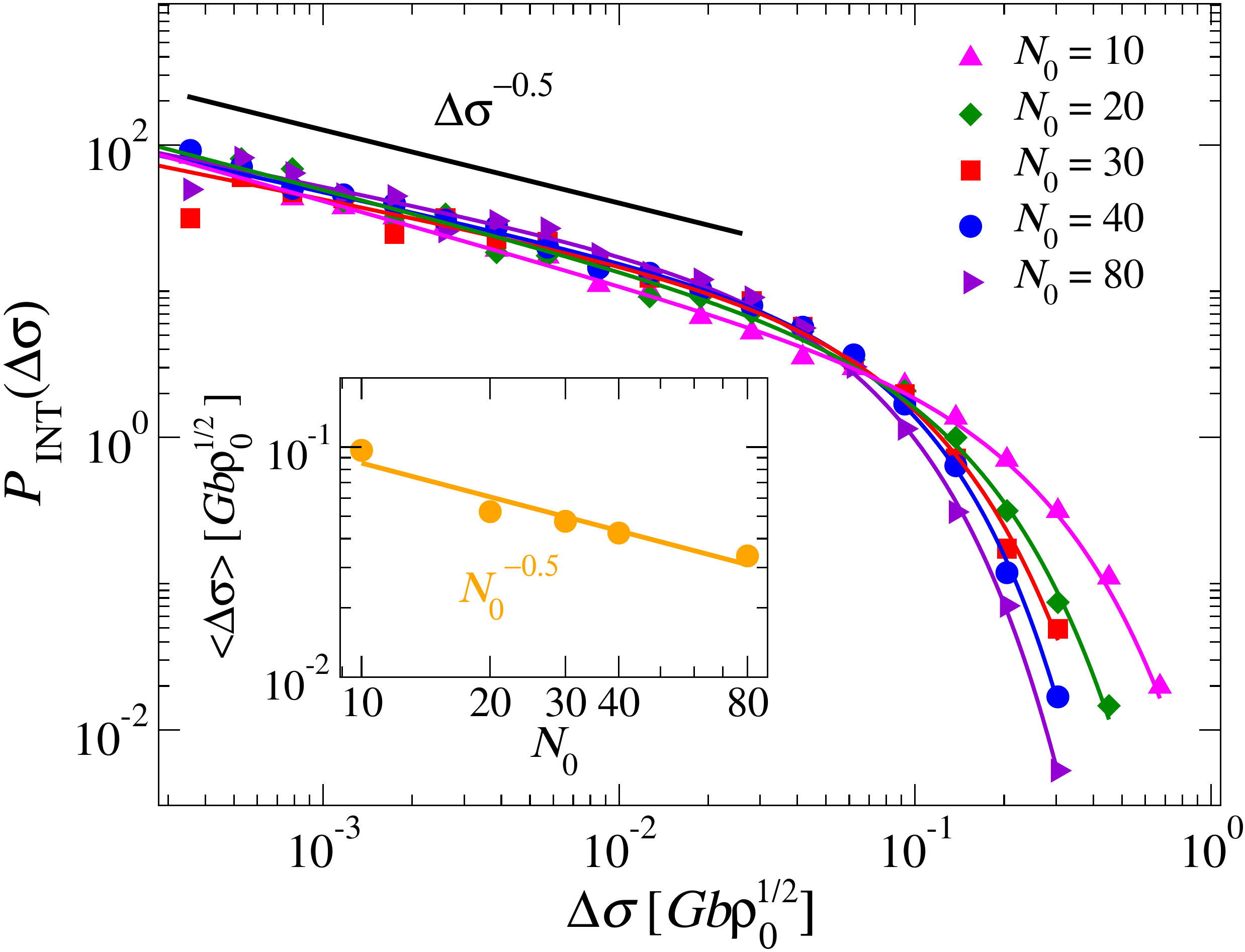}
\caption{(color online) Main figure shows the strain-integrated stress increment distributions
$P_\text{INT}(\Delta \sigma)$ for different $N_0$, revealing a cut-off decreasing with
the system size. The inset shows that the average stress jump magnitude $\langle 
\Delta \sigma \rangle$ decreases as $N_0^{-0.5}$.}
\label{fig:4}
\end{figure}

To further characterize the stress-dependence of the avalanche sizes, 
we show the scaling of the average total dislocation activity $\langle s \rangle D_\text{tot}$
vs $\sigma$ (with $s$ computed from the average velocity, and $D_\text{tot}=\sum_i l_i$ 
the total dislocation line length of the system) in the inset of Fig. \ref{fig:3}. 
We observe that the avalanche size increases roughly exponentially with stress for all 
system sizes $N_0$ and at any given stress it depends significantly on $N_0$. This is in 
contrast to a standard depinning transition where the avalanche size is independent on $N_0$ 
unless the stress is close to the depinning point. Similar results are, however, obtained in 
simplified 2D DDD models \cite{ISP-14,JAN-15} and experiments \cite{NG-08}. 
Our results indicate that rather than the applied stress, the limiting 
factor for the amount of dislocation activity within the strain bursts is the finite system 
size \cite{WEI-07}. Thus, the system appears to exhibit an extended, critical-like phase, with 
power-law distributed avalanches {\it at any applied stress}. This is in strong contrast to
tuned criticality observed in depinning-like non-equilibrium phase transitions where
criticality is observed only close to a critical point, and is analogous to glassy systems 
where similar extended critical phases have been observed \cite{BAK-07,MUL-15,PAZ-99}. 
Thus,``extended criticality'' seems to be a general 
feature of crystal plasticity of pure single crystals, irrespective of the spatial 
dimensionality of the system. Analogous ideas have very recently been presented also in the
context of amorphous plasticity \cite{LIN-15}.

The final issue we address concerns the statistics of stress increments $\Delta \sigma$, 
i.e. the vertical segments in the top inset of Fig. \ref{fig:1} (a); it is another 
quantity encoding information about the nature of the deformation process \cite{LIN-14}. 
Fig. \ref{fig:4} shows the $P_\text{INT}(\Delta \sigma)$ distributions of all stress 
increments along the stress-strain curves separating strain bursts larger than 
$s'^*$. These are power-law distributed up to a $N_0$ depedendent cutoff. The average 
stress increment $\langle \Delta \sigma \rangle$ decreases with $N_0$ as 
$\langle \Delta \sigma \rangle \propto N_0^{-0.6}$ (Fig. \ref{fig:4}, inset). A 
similar power law dependence of stress increments on the system size is measured 
experimentally in molybdenum micropillars \cite{ZAI-08}.

\section{Summary}

To conclude, we have shown that bursty three-dimensional crystal plasticity cannot be 
envisaged in terms of a depinning transition, but is rather a manifestation of an extended 
critical-like phase, reminiscent of glassy systems \cite{BAK-07,MUL-15,PAZ-99}. Interesting 
extensions of our study could be performed by adding a significant population of pinning 
centres, representing the effect of various additional defects such as precipitates \cite{LEH-16}, 
acting as obstacles for dislocation motion. Recent 2D studies \cite{OVA-15} suggest that 
when in the competition between dislocation jamming and pinning due to obstacles the 
latter starts to dominate, a depinning-like scenario may be recovered. Our results point 
out to the possibility that there are several universality classes in mesoscopic plasticity 
starting from the pure case studied here. Thus, the possible role of e.g. the crystal 
structure (FCC vs BCC, etc. \cite{WEI-15,BIS-15}) in determining the dislocation avalanche 
statistics in mesoscale 3D plasticity should be addressed. Our results await in-depth 
experimental studies.

\begin{acknowledgments}
AL, LL and MJA  are supported by the Academy of Finland through projects 13260053 and  251748 
(Centres of Excellence Programme, 2012-2017) and acknowledge the 
computational resources provided by the Aalto University School of Science ``Science-IT'' 
project, as well as those provided by CSC (Finland). LL is supported by an Academy
Research Fellowship (project no. 268302). SZ acknowledges support from the Academy of Finland 
FiDiPro progam, project 13282993. GC and SZ are supported by the European Research 
Council Advanced Grant n. 291002 SIZEFFECTS. 

\end{acknowledgments}

\end{document}